\documentclass[draftclsnofoot,onecolumn]{IEEEtran}
\IEEEoverridecommandlockouts

\usepackage{graphicx}
\usepackage{caption}
\usepackage{amsmath,graphicx,amssymb,mathtools,bm}
\usepackage{subfigure}
\usepackage{hyperref}
\usepackage{cite}
\usepackage{amsmath,amssymb,amsfonts}
\usepackage{algorithmic}

\usepackage{textcomp}
\usepackage{xcolor}
\usepackage{verbatim}  
\usepackage{graphicx}  
\usepackage{bm}  
\usepackage{mathrsfs} 
\usepackage{algorithm} 
\usepackage{algorithmic} 
\usepackage{booktabs}
\usepackage{textcomp}  
\usepackage{multirow}  
\usepackage{lettrine}   
\usepackage{graphicx}  
\usepackage{color}  
\usepackage{amsmath}
\usepackage{amssymb}

\usepackage{float} 

\def\BibTeX{{\rm B\kern-.05em{\sc i\kern-.025em b}\kern-.08em
    T\kern-.1667em\lower.7ex\hbox{E}\kern-.125emX}}
\begin{document}
	\newcommand{\tabincell}[2]{\begin{tabular}{@{}#1@{}}#2\end{tabular}} 

\title{Low-complexity Sparse Array Synthesis Based on Off-grid Compressive Sensing
}

\author{Songjie Yang,
	Baojuan Liu,
	Zhiqin Hong,
	Zhongpei Zhang,~\IEEEmembership{Member,~IEEE},
	
	\thanks{This work was supported in part by the National Key Research and Development Program of China under Grant 2020YFB1806800. (\textit{Corresponding author:
			Zhongpei~Zhang}.)
	}
	
	\thanks{Songjie Yang, Baojuan Liu, Zhiqin Hong and Zhongpei Zhang are with the National Key Laboratory of Science and Technology on Communications, University of Electronic Science and Technology of China, Chengdu 611731, China. (e-mail:
		yangsongjie@std.uestc.edu.cn; baojuanl@yeah.net ;202021220520@std.uestc.edu.cn; zhangzp@uestc.edu.cn).

	}
}
\maketitle

\begin{abstract}
In this letter, a novel sparse array synthesis method for non-uniform planar arrays is proposed, which belongs to compressive sensing (CS)-based systhesis.
Particularly, we propose an off-grid refinement technique to simultaneously optimize the antenna element positions and excitations with a low complexity, in response to the antenna position optimization problem that is difficult for standard CS. More importantly, we take into account the minimum inter-element spacing constraint for ensuring the physically realizable solution. Specifically, the off-grid Orthogonal Match Pursuit (OMP) algorithm is first proposed with low complexity and then off-grid Look Ahead Orthogonal Match Pursuit (LAOMP) is designed with better synthesis performance but higher complexity. In addition, simulation results have shown the proposed schemes have more advantages in computational complexity and synthesis performances compared with the related method.

\end{abstract}
\begin{IEEEkeywords}
Sparse array synthesis, off-grid compressive sensing, inter-element spacing constraint.
\end{IEEEkeywords}
\section{Introduction}
\lettrine[lines=2]{I}n order to reduce the cost of large-scale arrays such as massive multiple-input multiple-output (MIMO) communications, it is of paramount significance to reduce the number of antennas while maintain system performances as much as possible. In recent years, several studies on sparse array synthesis have tried to achieve certain pattern definitions with the smallest amount of antenna elements by optimizing their respective excitation coefficients and positions. 

In general, literature on sparse arrays can be broadly divided into several categories. The first is global optimization, which converts the synthesis problem into a binary optimization problem by selecting some elements on and others off, followed by solving it with optimization algorithms \cite{sparse1,sparse2,sparse3,sparse4,sparse5,sparse6,sparse7}, with a focus on genetic algorithms, generalized Gaussian quadrature approaches and particle swarm optimization, etc. However, this kind of method requires high comuputational complexity. The second method is Matrix pencil method (MPM), which can be used in the synthesis of non-uniform arrays \cite{MPM1,MPM2}. Thanks to the emergence of the compressed sensing (CS) paradigm \cite{CS}, the design of sparse arrays has vastly improved. Some CS techniques have been adopted to find the sparsest solution of the array synthesis, e.g., iterative weighted $L_1$ norm \cite{CS1,CS7,L1,IEDA}, two-step convex optimization \cite{CS2} and Bayesian compressive sensing (BCS) \cite{CS3,CS4,CS5,BCS4}. In BCS, the desired pattern is achieved via seeking the maximally-sparse array with the highest \emph{a-posteriori} probability. To synthesize the sparse array with non-uniform layout to obtain excellent synthesis performance, some of these CS-based schemes require the initial setting of dense elements.
 However, such a massive antenna placement exists two drawbacks. One is that it may lead to too small inter-element spacing in the sparse solution, which will incur the strong mutual coupling effect in the practical system, another shortcoming is the huge computational complexity required when applied to large-scale arrays. Thus, therein lies the challenge in the non-uniform array synthesis: making the optimized elements positions meet the minimum spacing constraint, so as to become practically realizable. For addressing this problem, \cite{CS6} proposed a constrained iterative weighted $L_1$ norm to find the sparse solution satisfying the minimum inter-element spacing. More recently, the authors in \cite{LAOMP1} exploited the property that the look ahead orthogonal matching pursuit (LAOMP) algorithm can select multiple atoms in each iteration to find the optimal atom that conforms the minimum inter-element spacing. However,
these approaches need to conduct CS algorithms in a dense antenna array, incurring considerable computational complexity. Due to constrained isometry (RIP) \cite{RIP}, dense element settings may not be optimal.

Considering the above, this letter is aimed at proposing an innovative CS-based synthesis scheme of non-uniform planar arrays with low complexity, where an off-grid CS framework with the minimum inter-element spacing constraint is adopted. 
Different from the mentioned CS strategies in \cite{CS1,CS7,CS2,L1,IEDA,CS3,CS4,CS5,BCS4,CS6,LAOMP1} with a candidate dense position grid, our proposed scheme optimize elements' positions and excitations via a iterative machanism, enabling better synthesis performance and lower computational complexity. In summary, our work presents an off-grid refinement-based sparse planar array synthesis technique that takes into account the minimal inter-element spacing constraint and computing cost.

\emph{Notations:} $\mathbb{C}^{x \times y}$ represents the complex-value matrices with the space of $ x \times y $. $(\cdot)^T$, $(\cdot)^*$ and $(\cdot)^H$ denote transpose, conjugate and conjugate transpose, respectively. $\vert\cdot\vert$, $\Vert \cdot\Vert_0$ and $\Vert \cdot\Vert_2$ are modulus, $\ell_0$ norm and $\ell_2$ norm, respectively.  Finally, $[\mathbf{b}]_i$ denotes the $i$-th element of vector $\mathbf{b}$.

\section{Problem Formulation}\label{PF}
Postulate that an $M\times N$ uniform planar array (UPA) lies on the $x$-$y$ plane. The element positions are $\{(x_m,y_n) | m=1,2,\cdots,M, \ n=1,2,\cdots,N\}$.
The corresponding set of excitations is $\{w_{mn} | m=1,2,\cdots,M, \ n=1,2,\cdots,N\}$. Thus, the array radiation pattern of the UPA is written as
\begin{equation}\label{F1}
	\mathcal{S}(u,v)=\sum_{m=1}^{M}\sum_{n=1}^{N}w_{mn} e^{{j\frac{2\pi}{\lambda}(x_mu+y_nv)}},
\end{equation}
where $\lambda$ is the antenna wavelength, $u={\rm sin}(\theta){\rm cos}(\varphi)$, $v={\rm sin}(\theta){\rm sin}(\varphi)$, $\theta$ and $\varphi$ are the sets of observed elevation and azimuth angles respectively. Suppose that there are $P$ elevation angles and $Q$ azimuth angles over the whole observation space, i.e., total $PQ$ observation points. Then we have
\begin{equation}\label{F2}
	\mathcal{S}(\bm{u},\bm{v})=\mathbf{A}\mathbf{w},
\end{equation}
where $\mathbf{A}\in\mathbb{C}^{PQ\times MN}$ and $\mathbf{w}\in\mathbb{C}^{MN\times 1}$ follow
\begin{equation}
	\begin{aligned}
	\mathbf{A}= &
\begin{bmatrix}
	e^{{j\frac{2\pi}{\lambda}(x_1 u_1+y_1v_1)}} & \cdots & e^{{j\frac{2\pi}{\lambda}(x_M u_1+y_Nv_1)}} \\
	\vdots  & \ddots & \vdots \\
	e^{{j\frac{2\pi}{\lambda}(x_1 u_{PQ}+y_1v_{PQ})}} & \cdots & e^{{j\frac{2\pi}{\lambda}(x_M u_{PQ}+y_Nv_{PQ})}}
\end{bmatrix} \\
=&[\mathbf{a}(x_1,y_1),\cdots,\mathbf{a}(x_M,y_N)]
	\end{aligned}
\end{equation}
and $
	\mathbf{w}=[w_{11},w_{12},\cdots,w_{MN}]^T,
$
where
\begin{equation}
	\mathbf{a}(x_m,y_n)=[e^{{j\frac{2\pi}{\lambda}(x_m u_1+y_nv_1)}},\cdots,e^{{j\frac{2\pi}{\lambda}(x_m u_{PQ}+y_nv_{PQ})}}]^T.
\end{equation} 

Therefore, the sparse array synthesis problem of the UPA can be describled as 
\begin{equation}\label{F}
	\begin{aligned}
		&  \underset{\mathbf{w}}{\rm arg \ min} \ \Vert \mathbf{w}\Vert_0\\
		{\rm s.t.} \ &\Vert \mathcal{S}_d(\bm{u},\bm{v})-\mathbf{A w}\Vert_2^2 \leq \varepsilon,
	\end{aligned}
\end{equation}
where $\mathcal{S}_d(\bm{u},\bm{v})\in\mathbb{C}^{PQ\times 1}$ is the desired radiation pattern and $\varepsilon$ is the matching accuracy parameter. As shown in x,

\section{Off-Grid CS with inter-element spacing constraints}\label{off-grid}
The methods to solve (\ref{F}) can be roughly divided into several categories: convex optimization, greedy algorithm and Bayesian learning. Generally, convex optimization and Bayesian learning incur high computational complexity when faced with large-scale antenna arrays.
 For computational efficiency, we adopt the orthogonal match pursuit (OMP) algorithm and propose an off-grid OMP for element position optimization with inter-element spacing constraints.
\begin{algorithm}[!t] 
	\caption{Off-Grid OMP With Antenna Spacing Constraints} 
	\label{off-omp}      
	\begin{algorithmic}[1] 
		\footnotesize{
			\REQUIRE {Desired pattern $\mathcal{S}_d(\bm{u,v})$, sensing matrix $\mathbf{A}$, minimum inter-element spacing $D_c$, learning rate $\kappa$ and match precise $\varepsilon$. }
			\ENSURE {Synthesized $\hat{\mathbf{x}}$, $\hat{\mathbf{y}}$ and $\hat{\mathbf{w}}$.} 
			
			\STATE{$\textbf{Initialize:}$ Resdual $\mathbf{r}^0=\mathcal{S}_d(\bm{u,v})$, solution $\mathbf{w}^0=0$, solution support $\Lambda^0=\emptyset$  and $t=0$.
			}

			\REPEAT
			\STATE{ \emph{Match:} $\mathbf{z}^t=\mathbf{A}^H\mathbf{r}^t$.}
			\STATE{\emph{Update Support:} $\Lambda^{t+1}=\Lambda^t\cup\{{\underset{p}{\rm arg \ max}}\vert [\mathbf
				{z}^t]_p\vert\}$.}
			\STATE{\emph{Update Residual:}
				$\mathbf{w}^{t+1}=\underset{\mathbf{f}:{\rm supp}(\mathbf{f}\subseteq\Lambda^{t+1})}{\rm arg \ min} \Vert\mathbf{r}^0-\mathbf{Af}\Vert^2_2$ \\
				\ \ \ \ \ \ \ \ \ \ \ \ \ \ \ \ \ \ \ \ \ $\mathbf{r}^{t+1}=\mathbf{r}^t-\mathbf{A}\mathbf{w}^{t+1}$ \\
				\ \ \ \ \ \ \ \ \ \ \ \ \ \ \ \ \ \ \ \ \
				$t=t+1$.
			}
			\STATE{ \textbf{\%\% Optimize element positions}  
			}
		\STATE{Construct ${\hat{\mathbf{w}}}$, $\hat{\mathbf{x}}$ and  $\hat{\mathbf{y}}$ via extracting the $t$ non-zero elements of the recovered $N$-dimensional  vector $\mathbf{z}^t$}.
			\FOR {$i=1, 2, \cdots$, $\mathcal{J}$}
			
			\STATE{\emph{Renew} $\mathbf{P}_i^x$ and $\mathbf{P}_i^y$ according to Eqn. (\ref{PXY});}
			\STATE{	$\hat{x}_k^{i+1}$ =
				${\rm max}\{\hat{x}^{i}_{k+1}-\delta,{\rm min}\{\hat{x}^i_{k-1}+\delta,\hat{x}_k^i+\kappa\textbf{Re}\{\mathbf{P}_i^{\hat{x},\dagger}\mathbf{r}_i\}\hat{x}_k^i\} \}$;}
			\STATE{$\hat{y}_k^{i+1}$=
			$	{\rm max}\{\hat{y}^{i}_{k+1}-\delta,{\rm min}\{\hat{y}^i_{k-1}+\delta,\hat{y}_k^i+\kappa\textbf{Re}\{\mathbf{P}_i^{\hat{y},\dagger}\mathbf{r}_i\}\hat{y}_k^i\}\}$;}
		\STATE{ $\bm{\Phi}_{i+1}=[\mathbf{a}(\hat{x}_1^{i+1},\hat{y}^{i+1}_1),\cdots,\mathbf{a}(\hat{x}_t^{i+1},\hat{y}^{i+1}_t)]$;}
		
			,

		\STATE{
			$	\hat{\mathbf{w}}^{i+1}=(\bm{\Phi}_{i+1}^H\bm{\Phi}_{i+1})^{-1}\bm{\Phi}_{i+1}^H\mathcal{S}_d(\bm{u},\bm{v})$.
		}
			
			\ENDFOR

			\UNTIL{$\Vert \mathcal{S}_d(\bm{u},\bm{v})-\sum_{k=1}^{t}\hat{w}_k\mathbf{a}(\hat{x}_k,\hat{y}_k)\Vert_2^2\leq \epsilon$.}
			
		}
	\end{algorithmic}
\end{algorithm}

OMP is a greedy sparse recovery algorithm with low complexity, the brief account of which is as follows. In the $(t+1)$-th iteration, the OMP procedure first chooses the best atom according to the projection of residual $\mathbf{r}$ onto $\mathbf{A}$, i.e.,
\begin{equation}
		\setlength{\abovedisplayskip}{2.7pt}
{\underset{p}{\rm arg \ max}} \ \vert [\mathbf
{z}^t]_p\vert
\end{equation}
with $
		\setlength{\abovedisplayskip}{2.7pt}
\mathbf{z}^t=\mathbf{A}^H\mathbf{r}^t,
$
where $\mathbf{r}^t$ denotes current residual and $\mathbf{z}^t$ is the projection vector.
Following that, OMP updates the next residual by eliminating the influence of the atom, where the least squares algorithm is performed in the subspace spanned by all collected atoms to attain the corresponding sparse coefficients. This is described as 
$
\mathbf{r}^{t+1}=\mathbf{r}^t-\mathbf{A}\mathbf{w}^{t+1}
$
 with
 \begin{equation}
	\setlength{\abovedisplayskip}{2.7pt}
\mathbf{w}^{t+1}=\underset{\mathbf{f}:{\rm supp}(\mathbf{f}\subseteq\Lambda^{t+1})}{\rm arg \ min}\Vert \mathcal{S}_d(\bm{u},\bm{v})-\mathbf{Af}\Vert_2^2,
	\setlength{\abovedisplayskip}{2.7pt}
 \end{equation}
where $\Lambda^{t+1}$ is the support set accommodating the selected $t+1$ atom indices, ${\rm supp}(\cdot)$ denotes the sparsity support, and $\mathbf{f}$ is the sparse vector.

In order to further optimize the positions of the antenna elements selected via the sparse recovery algorithm, so as to achieve better synthesis performance and ensure  practically realizable array architecture, we design a Taylor expansion-based OMP algorithm which iteratively optimizes the sparse element positions with the minimum inter-element spacing constraint considered. That is, the element position optimization problem can be carried out in each iteration of OMP and described as 
\begin{equation}\label{S}
		\setlength{\abovedisplayskip}{2.7pt}
	\begin{aligned}
	\underset{\eta_k,\zeta_k,\hat{w}_k}{\rm arg \ min}& \ \Vert \mathcal{S}_d(\bm{u},\bm{v})-\sum_{k=1}^{t}\hat{w}_k\mathbf{a}(\hat{x}_k+\eta_k\hat{x}_k,\hat{y}_k+\zeta_k\hat{y}_k)\Vert_2^2\\
	{\rm s.t.} &\ -\frac{\hat{x}_{k}-\hat{x}_{k-1}}{2}<\eta_k\hat{x}_k<\frac{\hat{x}_{k+1}-\hat{x}_k}{2}, \\ &-\frac{\hat{y}_{k}-\hat{y}_{k-1}}{2}<\zeta_k\hat{y}_k<\frac{\hat{y}_{k+1}-\hat{y}_k}{2}, 
	\end{aligned}
\end{equation}
where $t$ is the sparsity of recovered vector, i.e., the number of selected elements. $\hat{w}_k$, $(\hat{x}_k,\hat{y}_k)$ and $(\eta_k,\zeta_k)$ denote the excitation, position and perturbation of the $k$-th synthesized element. 

\begin{figure*}

	\begin{equation}\label{PXY}
		\begin{aligned}
			\mathbf{P}_i^x=&\left[
			\hat{w}_1\hat{x}_1\frac{\partial e^{{j\frac{2\pi}{\lambda}(\hat{x}_1 \bm{u}+\hat{y}_1\bm{v})}}}{\partial \hat{x}_1},\hat{w}_2\hat{x}_2\frac{\partial e^{{j\frac{2\pi}{\lambda}(\hat{x}_2 \bm{u}+\hat{y}_2\bm{v})}}}{\partial \hat{x}_2},\cdots,\hat{w}_t\hat{x}_t\frac{\partial e^{{j\frac{2\pi}{\lambda}(\hat{x}_t \bm{u}+\hat{y}_t\bm{v})}}}{\partial \hat{x}_t}\right],\\
			\mathbf{P}_i^y=&\left[
			\hat{w}_1\hat{y}_1\frac{\partial e^{{j\frac{2\pi}{\lambda}(\hat{x}_1 \bm{u}+\hat{y}_1\bm{v})}}}{\partial \hat{y}_1},\hat{w}_2\hat{y}_2\frac{\partial e^{{j\frac{2\pi}{\lambda}(\hat{x}_2 \bm{u}+\hat{y}_2\bm{v})}}}{\partial \hat{y}_2},\cdots,\hat{w}_t\hat{y}_t\frac{\partial e^{{j\frac{2\pi}{\lambda}(\hat{x}_t \bm{u}+\hat{y}_t\bm{v})}}}{\partial \hat{y}_t}\right],
		\end{aligned}
	\end{equation}
		\hrulefill
\end{figure*}
Note that the joint optimization in Eqn. (\ref{S}) is difficult to perform due to the coupling of variables. Hence, we tend to iteratively optimize $\eta_k,\zeta_k, \hat{w}_k,\forall k$. First of all, we approximate $\mathbf{a}(\hat{x}_k+\eta_k\hat{x}_k,\hat{y}_k+\zeta_k\hat{y}_k)$ via the second-order Taylor expansion with
\begin{equation}
		\setlength{\abovedisplayskip}{2.7pt}
	\begin{aligned}
\mathbf{a}&(\hat{x}_k+\eta_k\hat{x}_k,\hat{y}_k+\zeta_k\hat{y}_k)\approx\mathbf{a}(\hat{x}_k,\hat{y}_k)+\eta_k\hat{x}_k\frac{\partial\mathbf{a}(\hat{x}_k,\hat{y}_k)}{\partial \hat{x}_k}\\
&+\zeta_k\hat{y}_k\frac{\partial\mathbf{a}(\hat{x}_k,\hat{y}_k)}{\partial \hat{y}_k}+\frac{1}{2}(\eta_k\hat{x}_k\frac{\partial}{\partial \hat{y}_k}+\zeta_k\hat{y}_k\frac{\partial}{\partial \hat{y}_k})^2\mathbf{a}(\hat{x}_k,\hat{y}_k),
	\end{aligned}
\end{equation}
since the fourth term with respect to (w.r.t.) the square of perturbation $\{\eta_k,\zeta_k\}$
is small, it can be ignored.

 \begin{figure}
	\centering
	\label{chi_xi2}
	\includegraphics[width=0.4\textwidth]{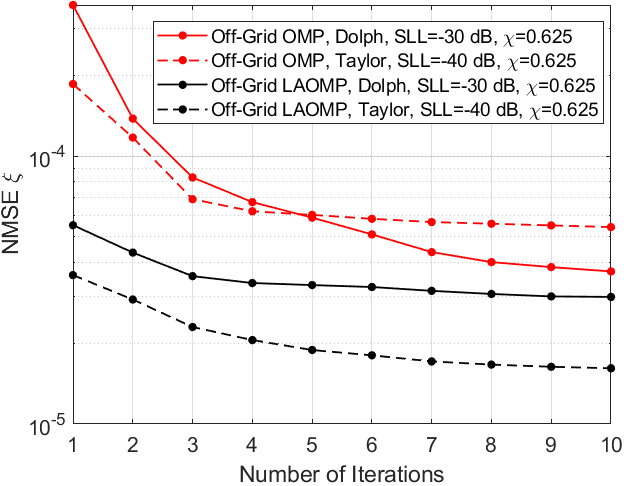}
	\caption{ 
		Pattern matching errors of proposed schemes versus the iteration number.}
\end{figure}
Along with this, the alternating algorithm is adopted to address Eqn. (\ref{S}). Given $\{\hat{w}_k\}_{k=1}^t$, we attain
\begin{equation}\label{SS}
		\setlength{\abovedisplayskip}{2.7pt}
\begin{aligned}
	\underset{\eta_k,\zeta_k}{\rm arg \ min} \ \Vert \mathbf{r}& - \sum_{k=1}^{t}\hat{w}_k\left( \eta_k\hat{x}_k\frac{\partial\mathbf{a}(\hat{x}_k,\hat{y}_k)}{\partial \hat{x}_k}+\zeta_k\hat{y}_k\frac{\partial\mathbf{a}(\hat{x}_k,\hat{y}_k)}{\partial \hat{y}_k}\right)\Vert_2^2\\
	{\rm s.t.} &\ -\frac{\hat{x}_{k}-\hat{x}_{k-1}}{2}<\eta_k\hat{x}_k<\frac{\hat{x}_{k+1}-\hat{x}_k}{2}, \\ &-\frac{\hat{y}_{k}-\hat{y}_{k-1}}{2}<\zeta_k\hat{y}_k<\frac{\hat{y}_{k+1}-\hat{y}_k}{2}, 
\end{aligned}
\end{equation}
where $\mathbf{r}=\mathcal{S}_d(\bm{u},\bm{v})-
\sum_{k=1}^{t}
\hat{w}_k\mathbf{a}(\hat{x}_k,\hat{y}_k)$ is the residual. Given iteration number $i$, the element positions of the next iteration can be updated sequentially via an alternating way. 
 Consider the constraint of minimum inter-element spacing $D_c$, we define $\delta=\frac{\sqrt{2}}{2}D_c$ to confine the optimization of $\hat{x}$ and $\hat{y}$,
 i.e.,
\begin{equation}
		\setlength{\abovedisplayskip}{2.7pt}
	\hat{x}_k^{i+1}=
		{\rm max}\{\hat{x}^{i}_{k+1}-\delta,{\rm min}\{\hat{x}^i_{k-1}+\delta,\hat{x}_k^i+\kappa\textbf{Re}\{\mathbf{P}_i^{\hat{x},\dagger}\mathbf{r}_i\}\hat{x}_k^i\} \},
\end{equation}
\begin{equation}
	\hat{y}_k^{i+1}={\rm max}\{\hat{y}^{i}_{k+1}-\delta,{\rm min}\{\hat{y}^i_{k-1}+\delta,\hat{y}_k^i+\kappa\textbf{Re}\{\mathbf{P}_i^{\hat{y},\dagger}\mathbf{r}_i\}\hat{y}_k^i\} \},
\end{equation}
where $\textbf{Re}\{\cdot\}$ denotes the real part, $\kappa$ is the learning rate, $\mathbf{P}_i^{x,\dagger}=(\mathbf{P}_i^{x,H}\mathbf{P}_i^{x})^{-1}\mathbf{P}_i^{x,H}$, $\mathbf{P}_i^{y,\dagger}=(\mathbf{P}_i^{y,H}\mathbf{P}_i^{y})^{-1}\mathbf{P}_i^{y,H}$, $\mathbf{P}_i^{x}\in\mathbb{C}^{PQ\times t}$ and $\mathbf{P}_i^y\in\mathbb{C}^{PQ\times t}$ are the partial matrices w.r.t. $\mathbf{x}^i$ and $\mathbf{y}^i$, respectively, as shown in Eqn. (\ref{PXY}) at the top of this page.

With $\{\hat{x}^{i+1}_k,\hat{y}^{i+1}_k\}_{k=1}^t$ attained, their corresponding excitation coefficients can be updated by 
\begin{equation}\label{W}
		\hat{w}^{i+1}_k=\underset{\hat{w}_k}{\rm arg \ min} \ \Vert \mathcal{S}_d(\bm{u},\bm{v})-\sum_{k=1}^{t}\hat{w}_k\mathbf{a}(x_k^{i+1},y^{i+1}_k)\Vert_2^2.
\end{equation}
This can be easily addressed by the least squares algorithm as follows.
\begin{equation}
		\setlength{\abovedisplayskip}{2.7pt}
	\hat{\mathbf{w}}^{i+1}=(\bm{\Phi}_{i+1}^H\bm{\Phi}_{i+1})^{-1}\bm{\Phi}_{i+1}^H\mathcal{S}_d(\bm{u},\bm{v}),
\end{equation}
where $\bm{\Phi}_{i+1}=[\mathbf{a}(\hat{x}_1^{i+1},\hat{y}^{i+1}_1),\cdots,\mathbf{a}(\hat{x}_t^{i+1},\hat{y}^{i+1}_t)]\in\mathbb{C}^{PQ\times t}$.

 In general, we assume that $\hat{\mathbf{w}}$, $\hat{\mathbf{x}}$ and $\hat{\mathbf{y}}$ are iteratively optimized $\mathcal{J}$ times. Notably,
  the minimun inter-element spacing constraint is taken into account in each iteration. In summary, the whole synthesis flow is illustrated in Algorithm \ref{off-omp}. Indeed, our proposed off-grid method can be easily extended to the LAOMP algorithm.
 
\section{Simulation Results}\label{SR}
 In this section, some experiments are carried out to evaluate synthesis performances of the proposed scheme from several indexes of array sparsity rate $\chi$, normalized matching error/normalized mean square error (NMSE) $\xi$ and sidelobe level (SLL), where the NMSE $\xi$ and the array sparsity rate $\chi$ are formulated as follows:
\begin{equation}\label{xi}
	\xi =\frac{\int_{-1}^{1}\int_{-1}^{1} \vert\mathcal{S}
		(u,v)-\mathcal{S}_d(u,v)\vert^2 du dv}{\int_{-1}^{1}\int_{-1}^{1} \vert\mathcal{S}_d(u,v)\vert^2 du dv},
\end{equation}
\begin{equation}
		\setlength{\abovedisplayskip}{2.7pt}
	\chi = \frac{t}{MN},
\end{equation}
 where $S(u,v)$ is the synthesized pattern, $MN$ represents the number of antennas of the reference pattern and $t$ is that of the achieved pattern.
 	 \begin{figure}
 		\centering
 		\includegraphics[width=0.4\textwidth]{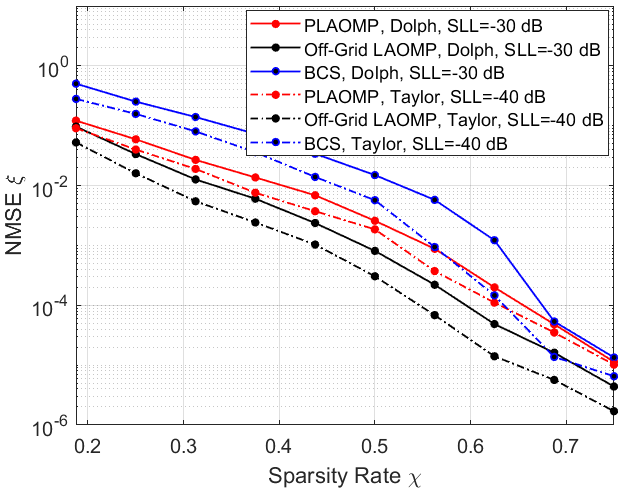}
 		\caption{Pattern matching errors of different schemes versus sparsity $\chi$.}
 	\end{figure}
 
 \begin{figure*}[htbp]
 	\centering
 	\subfigure[Dolph Reference, SLL=-30 dB ]{
 		\includegraphics[width=0.23\textwidth]{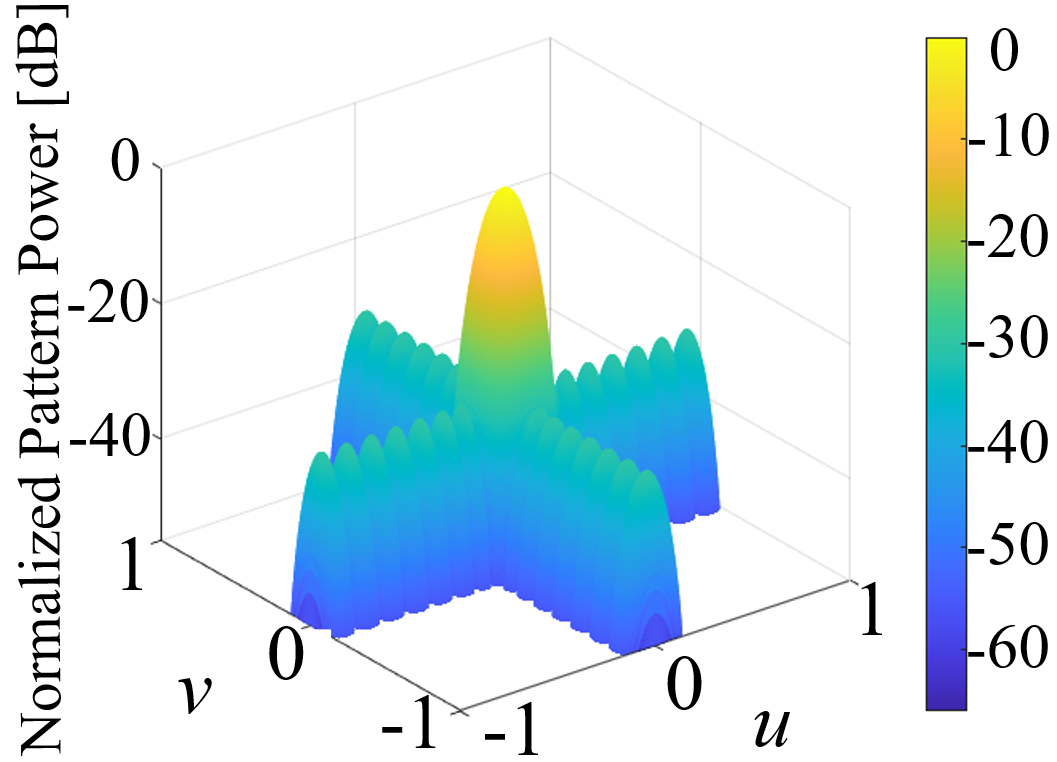}
 	}
 	\subfigure[Dolph Synthesis, SLL=-29.91 dB]{
 		\includegraphics[width=0.23\textwidth]{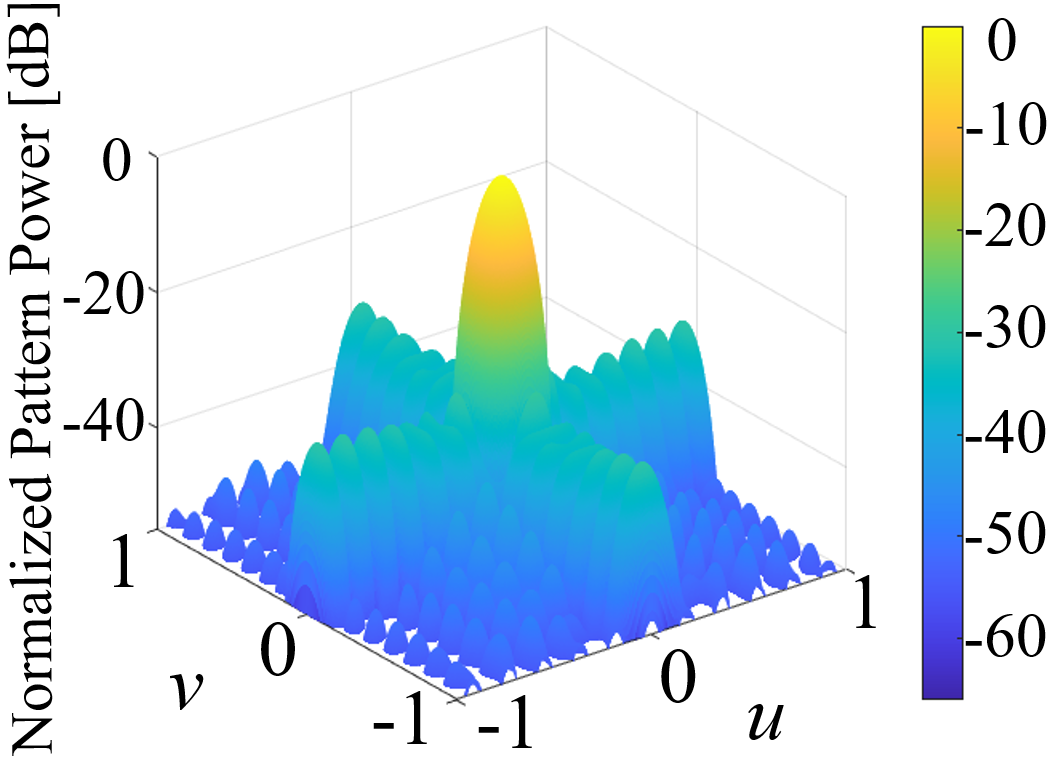}
 	}
 	\subfigure[Taylor Reference, SLL=-40 dB]{
 		\includegraphics[width=0.23\textwidth]{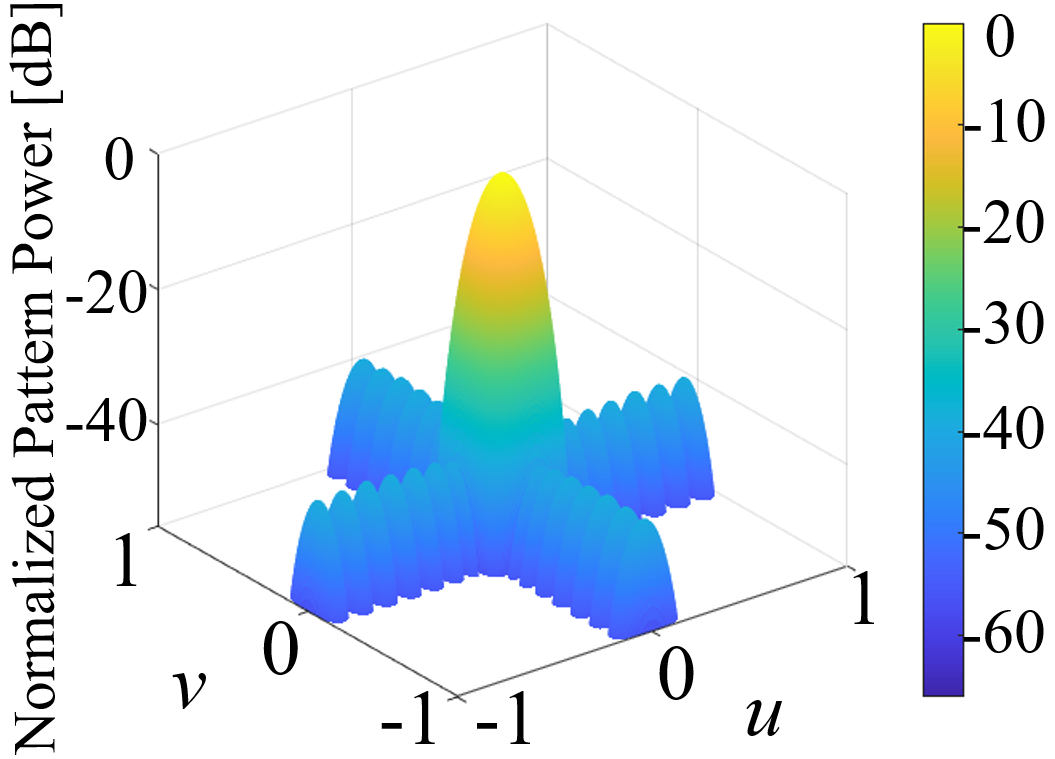}
 	}
 	\subfigure[Taylor Synthesis, SLL=-39.86 dB]{
 		\includegraphics[width=0.23\textwidth]{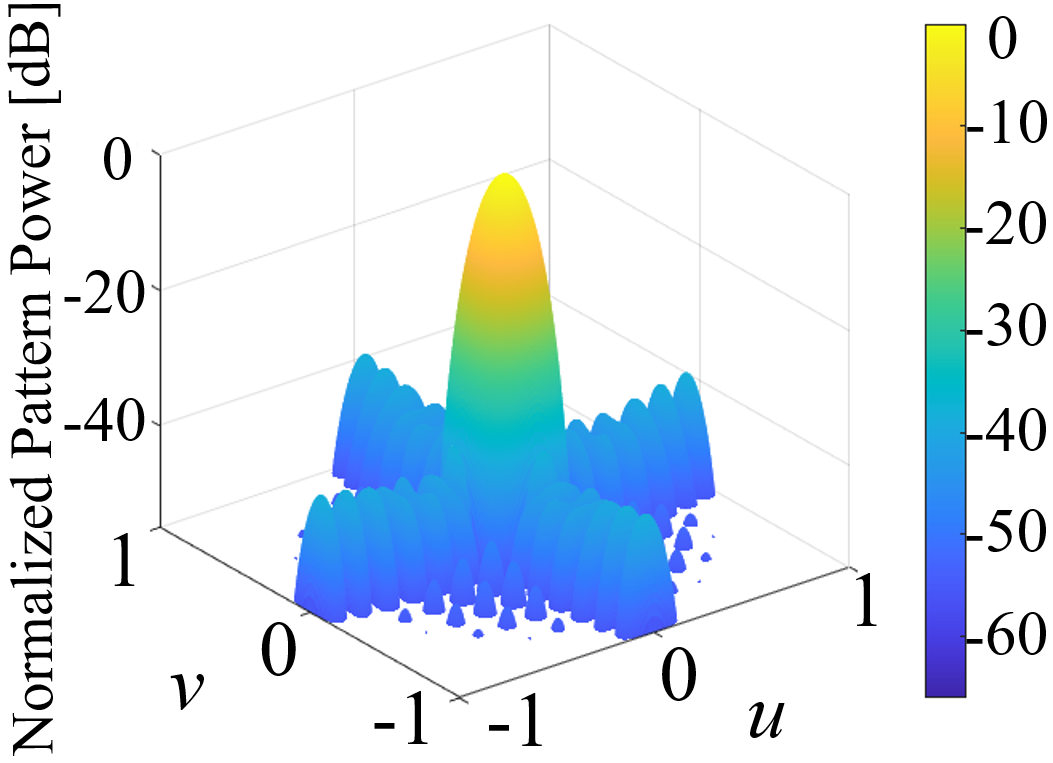}
 	}
 	\caption{Beam patterns of Dolph and Taylor reference and  synthesized patterns via our proposed off-grid LAOMP.}
 	\label{pattern}
 \end{figure*}
 
 \begin{figure*}[htbp]
 	\centering
 	\subfigure[Dolph Reference, $M\times N=256$]{
 		\includegraphics[width=0.23\textwidth]{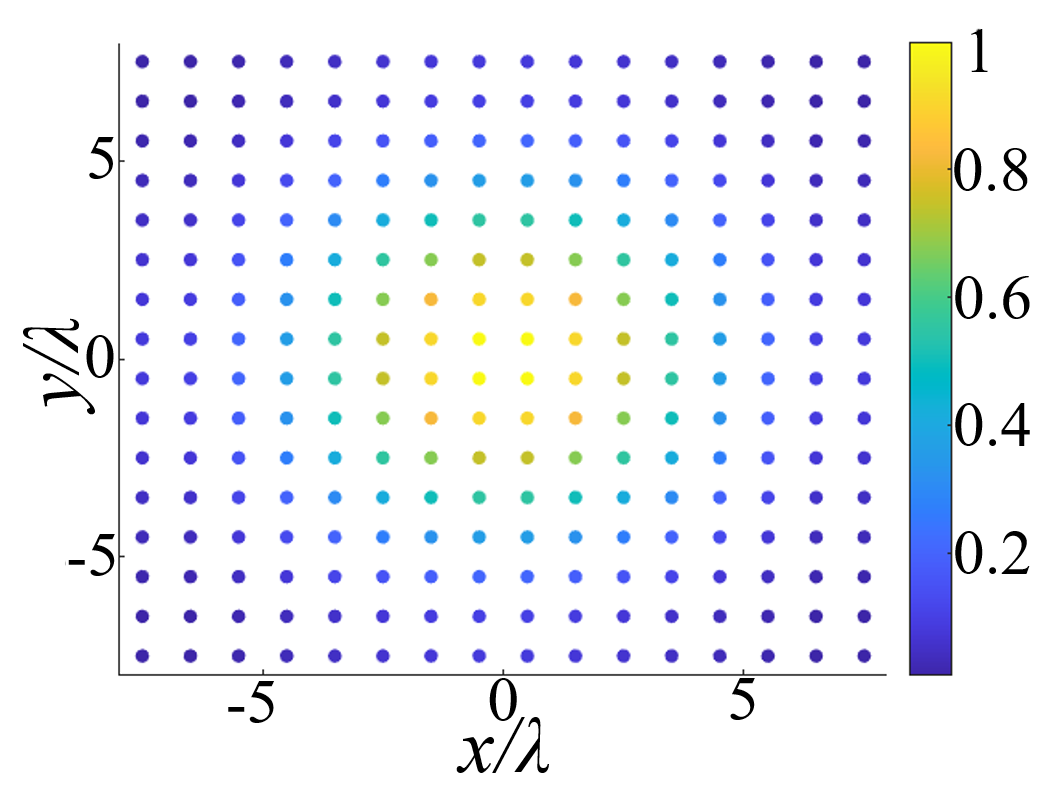}
 	}
 	\subfigure[Dolph Synthesis, $M\times N=160$]{
 		\includegraphics[width=0.23\textwidth]{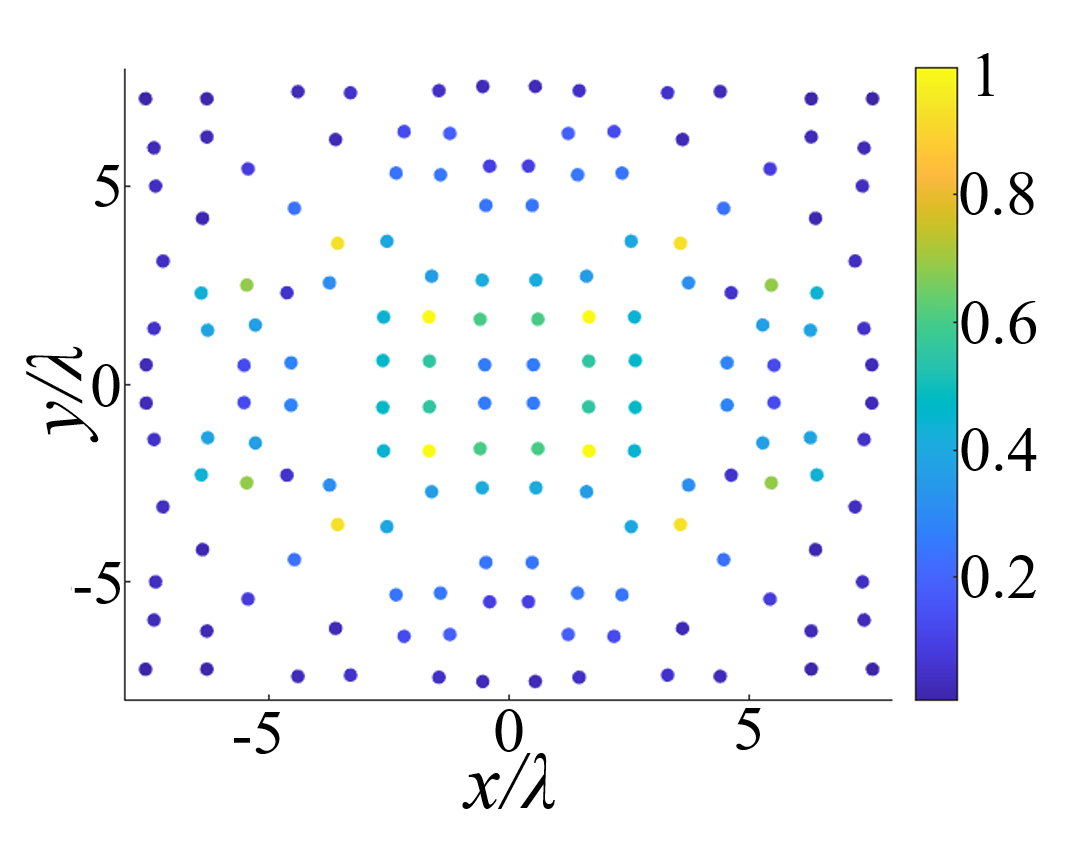}
 	}
 	\subfigure[Taylor Reference, $M\times N=256$]{
 		\includegraphics[width=0.23\textwidth]{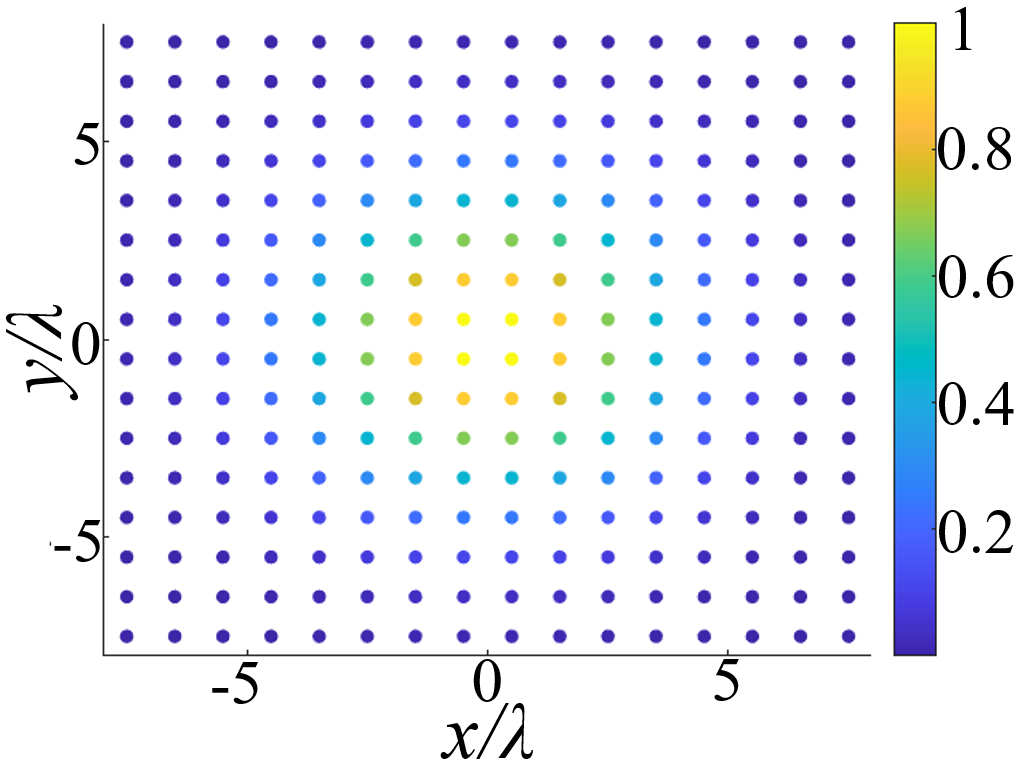}
 	}
 	\subfigure[Taylor Synthesis, $M\times N=160$]{
 		\includegraphics[width=0.23\textwidth]{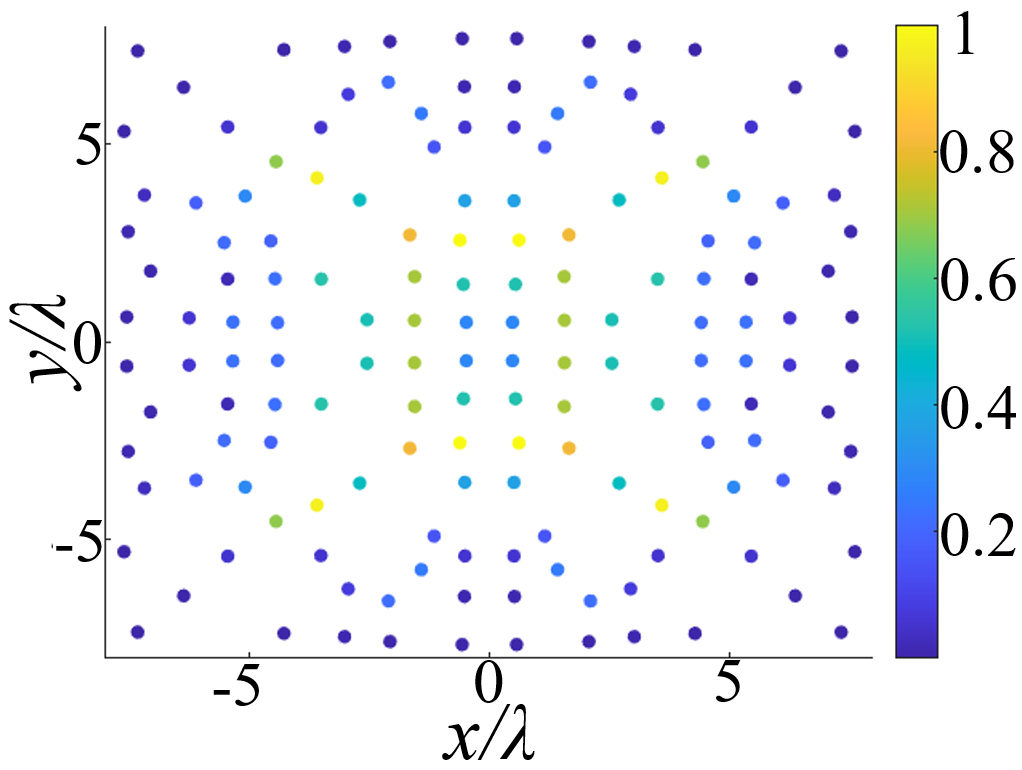}
 	}
 	\caption{Array element positions and normalized excitation coefficients of reference and synthesis patterns. }
 	\label{layout}
 \end{figure*}

 For the benchmark schemes, BCS \cite{CS5} and PLAOMP \cite{LAOMP1} with a dense grid are considered. In addition, we directly extend our proposed off-grid OMP to off-grid LAOMP for better performance but higher complexity.
 For the simulation setting, Dolph-Chebyscheff pattern with SLL$=-30$ dB and Taylor pattern with SLL$=-40$ dB are produced by a $16\times 16$ uniform planar array with half-wavelength spaced antennas. We set the minimum inter-element spacing $D_c$ to $0.5\lambda$ and the inter-element spacing of candidate antennas in PLAOMP and BCS to $0.01\lambda$ ($\Delta x=\Delta y=0.01\lambda$). Furthermore, the performance of LAOMP is affected by the look-ahead parameter which is set to $L=15$.

 The first experiment is to evaluate the impact of the iteration number $\mathcal{J}$ on the synthesis performances of proposed schemes shown in Fig. 1, where $t=160$, i.e., $\chi=0.625$ is set.
  This implies the synthesis performance starts to converge after several iterations. Without losing generality, we set $\mathcal{J}$ to 10 in consequent experiments.
In Fig. 2, we plot the two reference patterns' matching error versus the array sparsity rate, where $t$ ranges from 48 to 192, i.e., $\chi$ ranges from 0.1875 to 0.75. The maximal iteration number for the BCS algorithm and the look-ahead parameter for LAOMP-based algorithms are set to 1000 and 15, respectively. As can be seen, our proposed off-grid method outperforms PLAOMP and BCS in terms of the NMSE of the two patterns.

Next, we exhibit the reference patterns and synthesized patterns, where the parameters of reference patterns are as stated previously.  We use the proposed off-grid LAOMP synthesis method with $L=15$ and $\chi=0.625$. Fig. 3 illustrates the reference beam patterns and synthesized patterns. It's shown that the synthesized patterns are close to the reference patterns. In addition, the SLLs of the synthesized Dolph and Taylor patterns are -29.91 and -39.86 dB, respectively. Particularly, we visualize the element positions and excitation coefficients of the mentioned four patterns, as shown in Fig. 4.

\begin{table}
	\renewcommand\arraystretch{1.07} 
	\caption{A Comparison of Computational Complexity}
	\label{table1}
	\centering
	\begin{tabular}{ |c|c|}
		\hline
		{ Scheme}	& Computational Complexity \\ \hline\hline
		
		{BCS \cite{CS5}}		&  $\mathcal{O}(\mathcal{K}(G_x^3G_y^3+QPG_x^2G_y^2))$\\
		\hline
		{PLAOMP \cite{LAOMP1}}			&  $\mathcal{O}(\mathcal{J}_1\mathcal{J}_2LG_xG_yPQ)$ \\
		\hline
		{Proposed Off-Grid OMP}& $\mathcal{O}(\mathcal{L}MNPQ+\mathcal{L}^2\mathcal{J}PQ)$ \\
		\hline
	Proposed Off-Grid LAOMP& $\mathcal{O}(\mathcal{G}_1\mathcal{G}_2LMNPQ+\mathcal{G}_1^2\mathcal{J}PQ)$ 	
		\\		
		\hline
		
	\end{tabular}
\end{table}

Finally, the computational complexity of mentioned schemes is analyzed in Table \ref{table1}, where $\mathcal{K}$, $\mathcal{J}_1$, $\mathcal{J}_2$, $\mathcal{L}$, $\mathcal{G}_1$ and $\mathcal{G}_2$ are these algorithms' iteration numbers\footnote{LAOMP-based methods have two kinds of iterations, one is main loop and the other is the look-ahead procedure \cite{LAOMP1}.}. 
 First, BCS will incur a high level of complexity due to the inverse operation. The greatest difference between BCS/PLAOMP and our suggested methods is that the former relies on the grid density (i.e., $G_x$ and $G_y$). For instance, as we set $\Delta x=\Delta y=0.01\lambda$, $G_x=G_y=\frac{0.5\lambda}{0.01\lambda}\times MN=12800$. This will bring unacceptable computing resources, especially when $MN$ is large. However, our proposed schemes avoid high complexity since the initial elements are placed on an $M\times N$ grid. 
 In this case, further position optimization results in an additional complexity term produced by Algorithm \ref{off-omp} lines 9-13. In general, it is observable that the proposed schemes are faster than the benchmarks; for instance, off-grid LAOMP is about $\frac{G_xG_y}{MN}$ times faster than PLAOMP.

\section{Conclusions}\label{Con}
In this letter, we combine the off-grid concept with OMP and LAOMP to effectively synthesize the antenna element positions and excitations of sparse planar arrays, where the minimum inter-element spacing constraint is considered. Compared with the related method, BCS and PLAOMP, our proposed off-grid LAOMP achieves lower computational complexity and better synthesis performances. It is noteworthy that the proposed off-grid OMP can reduce the complexity by a large margin while ensure considerable performances when large-scale arrays are employed.

\bibliographystyle{IEEEtran}
\bibliography{reference.bib}

\vspace{12pt}

\end{document}